\documentclass{appolb}
\usepackage{amsmath,amsthm,amssymb,cite,enumerate}
\usepackage[a4paper,left=0.5in,right=1in]{geometry}
\usepackage[usenames]{color}
\usepackage{graphicx}
\usepackage{epsfig}
\usepackage{dcolumn}
\usepackage{bm}
\usepackage{authblk} 

\begin{document}

\def \d {{\rm d}}

\def \bF {\mbox{\boldmath{$F$}}}
\def \bV {\mbox{\boldmath{$V$}}}
\def \bff {\mbox{\boldmath{$f$}}}
\def \bT {\mbox{\boldmath{$T$}}}
\def \bk {\mbox{\boldmath{$k$}}}
\def \bl {\mbox{\boldmath{$\ell$}}}
\def \bn {\mbox{\boldmath{$n$}}}
\def \bbm {\mbox{\boldmath{$m$}}}
\def \tbbm {\mbox{\boldmath{$\bar m$}}}

\def \T {\bigtriangleup}
\newcommand{\msub}[2]{m^{(#1)}_{#2}}
\newcommand{\msup}[2]{m_{(#1)}^{#2}}

\newcommand{\be}{\begin{equation}}
\newcommand{\ee}{\end{equation}}

\newcommand{\beq}{\begin{eqnarray}}
\newcommand{\eeq}{\end{eqnarray}}
\newcommand{\pa}{\partial}
\newcommand{\pp}{{\it pp\,}-}
\newcommand{\ba}{\begin{array}}
\newcommand{\ea}{\end{array}}

\newcommand{\M}[3] {{\stackrel{#1}{M}}_{{#2}{#3}}}
\newcommand{\m}[3] {{\stackrel{\hspace{.3cm}#1}{m}}_{\!{#2}{#3}}\,}

\newcommand{\tr}{\textcolor{red}}
\newcommand{\tb}{\textcolor{blue}}
\newcommand{\tg}{\textcolor{green}}

\newcommand*\bg{\ensuremath{\boldsymbol{g}}}
\newcommand*\bE{\ensuremath{\boldsymbol{E}}}
\newcommand*\bh{\ensuremath{\boldsymbol{h}}}
\newcommand*\bR{\ensuremath{\boldsymbol{R}}}

\def\a{\alpha}
\def\b{\beta}
\def\g{\gamma}
\def\de{\delta}

\def\E{{\cal E}}
\def\B{{\cal B}}
\def\R{{\cal R}}
\def\F{{\cal F}}
\def\L{{\cal L}}

\def\e{e}
\def\bb{b}

\newtheorem{theorem}{Theorem}[section] 
\newtheorem{cor}[theorem]{Corollary} 
\newtheorem{lemma}[theorem]{Lemma} 
\newtheorem{prop}[theorem]{Proposition}
\newtheorem{definition}[theorem]{Definition}
\newtheorem{remark}[theorem]{Remark}  
\newtheorem{proposition}[theorem]{Proposition}

\title{On universal black holes\thanks{Presented at the 6th Conference of the Polish Society on Relativity, Szczecin, Poland, September 23--26, 2019}
}

\author[1]{Sigbj\o rn Hervik\thanks{sigbjorn.hervik@uis.no}}
\author[2]{Marcello Ortaggio\thanks{ortaggio(at)math(dot)cas(dot)cz}}

\affil[1]{Faculty of Science and Technology, University of Stavanger, N-4036 Stavanger, Norway}
\affil[2]{Institute of Mathematics of the Czech Academy of Sciences, \newline \v Zitn\' a 25, 115 67 Prague 1, Czech Republic}

\maketitle

\begin{abstract}
Recent results \cite{HerOrt19} on universal black holes in $d$ dimensions are summarized. These are static metrics with an isotropy-irreducible homogeneous base space which can be consistently employed to construct solutions to virtually any metric theory of gravity in vacuum.
\end{abstract}


\section{Introduction}

\label{intro}

Let us consider the static black hole ansatz
\beq
	\bg=e^{a(r)}\left(-f(r)\d t^2+\frac{\d r^2}{f(r)}\right) + r^2h_{ij}(x^k)\d x^i\d x^j .
	\label{metric1}
\eeq
When $a=0$, $f=1-\frac{\mu}{r}$ and $\bh=h_{ij}\d x^i\d x^j$ is the metric of a $2$-dimensional round unit sphere, this represents the well-known spherical Schwarzschild black hole of four-dimensional general relativity.

Extensions to Einstein's gravity in $d=n+2$ spacetime dimensions with a cosmological constant are readily obtained if one takes $f=K-\frac{\mu}{r^{d-3}}-\lambda r^2$ and $\bh$ is the metric of an $n$-dimensional Einstein space with Ricci scalar $\tilde R=n(n-1)K$ \cite{Tangherlini63,GibWil87,Birmingham99}. While $\bh$ can be {\em any} Einstein space in Einstein's gravity, obstructions to the permitted geometries arise in more general higher dimensional theories such as Gauss-Bonnet and Lovelock gravity \cite{DotGle05,FarDeh14,Ray15,OhaNoz15}. 

In our recent work \cite{HerOrt19}, we have studied the metric ansatz~\eqref{metric1} in higher-order vacuum gravity theories of the form
\be
 S=\int\d^dx\sqrt{-g}{\cal L}(\bR,\nabla\bR,\ldots) ,
\label{action}
\ee
where ${\cal L}$ is a scalar invariant constructed polynomially from the Riemann tensor $\bR$ and its covariant derivatives of arbitrary order. We have obtained a sufficient condition on the metric $\bh$ which enables the ansatz \eqref{metric1} to be consistently employed in any such theory, as we summarize in the following.

\section{Black Holes with universal horizons}

\label{sec_universal}
 
First of all, let us recall the following geometric definition (quoted, for convenience, from \cite{Bleecker79}):
\begin{definition}[IHS space]
\label{def_IHS}
 An {\em isotropy-irreducible homogeneous space} (IHS) $(M,\bh)$ is a homogeneous space whose isotropy group at a point acts irreducibly on the tangent space of $M$ at that point. 
\end{definition}

For our purposes, it is important to observe that an IHS is necessarily Einstein (but not vice versa) and, more generally,  for an IHS any symmetric 2-tensor on $M$ possessing the symmetries of $\bh$ must be proportional to $\bh$ \cite{Wolf68}. IHS are equivalent to {\em universal} Riemannian spaces in the sense explained in \cite{HerOrt19}. Examples of IHS can be found in \cite{Bessebook}. These include direct products of (identical) spaces of constant curvature and irreducible symmetric spaces. In particular, in $n=4$ dimensions, an IHS must symmetric and therefore locally one of the following: $S^4$, $S^2\times S^2$, $H^4$, $H^2\times H^2$, $\mathbb{C}P^2$, $H_{\mathbb{C}}^2$, or flat space (cf., e.g., \cite{Bessebook} and references therein). 

Now we can quote the main result of \cite{HerOrt19}:
\begin{prop}
 \label{prop_E}
Consider any metric of the form \eqref{metric1} where $\bh$ is an IHS. Then, any symmetric 2-tensor $\bE$ constructed from tensor products, sums and contractions from the metric $\bg$, the Riemann tensor $\bR$, and its covariant derivatives necessarily takes the form
\be  
	\bE=F(r)\d t^2+G(r) \d r^2+H(r) h_{ij}(x^k)\d x^i\d x^j . 
	\label{E}
\ee 
\end{prop}

Let us now note that the field equations derived from \eqref{action} (neglecting boundary terms) are of the form $\bE=0$, where $\bE$ is a symmetric, conserved rank-2 tensor locally constructed out of $\bg$ and its derivatives \cite{Eddington_book} (cf. also \cite{IyeWal94}). We can thus apply proposition~\ref{prop_E} to observe that, in any theory of gravity \eqref{action}, the tensorial field equation $\bE=0$ for the metric \eqref{metric1} with $\bh$ IHS reduces to three ``scalar'' equations $F(r)=0$, $G(r)=0$ and $H(r)=0$. Furthermore, the equation $H(r)=0$ holds automatically once $F(r)=0=G(r)$ are satisfied, thanks to the fact that $\bE$ is identically conserved. One is thus left with just {\em two ODEs} for the two metric functions $a(r)$ and $f(r)$ (their precise form will depend on the particular gravity theory under consideration -- several examples can be found in \cite{HerOrt19} and references therein). This is a drastic simplification of the tensorial field equation $\bE=0$.
These spacetimes will generically describe static black holes -- we name them {\em universal black holes} because they possess a universal (IHS) horizon and because the construction described above works universally in any theory \eqref{action}. The details (including the precise form of $a(r)$ and $f(r)$) and physical properties of the solutions depend on the specific theory one is interested in. Since for $n=2,3$ an $n$-dimensional Einstein space is necessarily of constant curvature, this result is of interest for dimension $d\ge6$ (i.e., $n\ge4$). 

Some comments on the near-horizon geometries associated with extremal limits of the universal black holes described above can be found in \cite{HerOrt19} (see also \cite{Gurses92}).

\section{Examples}

\label{sec_Eins}

Here we illustrate the results of section~\ref{sec_universal} by giving explicit examples of black holes solutions in certain gravity theories of the form~\eqref{action}. Quantities with a tilde will refer to the transverse space geometry of $\bh$ (taken to be IHS), so that 
\be
 \tilde{R}_{ij}=(n-1)Kh_{ij} ,
 \label{def_K}
\ee
and thus $\tilde{R}=n(n-1)K$.

\subsection{Gauss-Bonnet gravity}

\label{sec_GB}

This theory is defined by the Lagrangian density 
\be
 {\cal L}=\sqrt{-g}\left[\frac{1}{\kappa}(R-2\Lambda)+\gamma I_{GB}\right] , \qquad I_{GB}=R_{\mu\nu\rho\sigma}R^{\mu\nu\rho\sigma}-4R_{\mu\nu}R^{\mu\nu}+R^2 ,
 \label{GB}
\ee
where $\kappa$, $\Lambda$ and $\gamma$ are constants. 

With the ansatz \eqref{metric1}, it possesses the black hole solution \cite{DotGle05,DotOliTro09,Bogdanosetal09,Maeda10,DotOliTro10}
\beq
 & & a(r)=0 , \\ 
 & & f(r)=K+\frac{r^2}{2\kappa\hat\gamma}\left[1\pm\sqrt{1+4\kappa\hat\gamma\left(\frac{2\Lambda}{n(n+1)}+\frac{\mu}{r^{n+1}}\right)-\frac{4\kappa^2\hat\gamma^2\tilde I_W^2}{r^4}}\right] ,
 \label{f_GB}
\eeq
where $\mu$ is an integration constant and
\be
 \hat\gamma=(n-1)(n-2)\gamma , \qquad n(n-1)(n-2)(n-3)\tilde I_W^2=\tilde C_{ijkl}\tilde C^{ijkl} .
 \label{IW}
\ee

Eq.~\eqref{f_GB} shows that the Weyl tensor of the geometry $\bh$ affects the solution. The branch with the minus sign admits a GR limit by taking $\hat\gamma\to0$. The non-negative constant $\tilde I_W^2$ vanishes iff $\bh$ is conformally flat (so necessarily when $n=3$), 
in which case one recovers the well-known black holes with a constant curvature base space \cite{BouDes85,Wheeler86_GB,Cai02}.

\subsection{Pure cubic Lovelock gravity}

\label{sec_Lov}

In more than six dimensions, a natural extension of Gauss-Bonnet (and Einstein) gravity is given by Lovelock gravity \cite{Lovelock71}. The special {\em purely cubic} theory is defined by
\be
 {\cal L}=\sqrt{-g}(c_0+c_3{\cal L}^{(3)}) , \qquad\qquad {\cal L}^{(3)}=\frac{1}{8}\delta_{\mu_1\nu_1\mu_2\nu_2\mu_3\nu_3}^{\rho_1\sigma_1\rho_2\sigma_2\rho_3\sigma_3}R_{\rho_1\sigma_1}^{\mu_1\nu_1}R_{\rho_2\sigma_2}^{\mu_2\nu_2}R_{\rho_3\sigma_3}^{\mu_3\nu_3} ,
 \label{Lagr}
\ee
$\delta^{\mu_1\ldots \mu_p}_{\rho_1\ldots \rho_p}=p!\delta^{\mu_1}_{[\rho_1}\ldots\delta^{\mu_p}_{\rho_p]}$ and $c_0$, $c_3$ are constants.

It possesses the solution
\be
 a(r)=0 ,
\ee
\beq
f(r)-K= & & \frac{1}{(2\hat c_3)^{1/3}}\left[c_0r^6-\frac{\mu}{r^{n-5}}+\hat c_3\tilde J_W+\sqrt{\left(c_0r^6-\frac{\mu}{r^{n-5}}+\hat c_3\tilde J_W\right)^2+4\hat c_3^2\tilde I_W^6}\right]^{1/3} \nonumber \\
			 & & {}+\frac{1}{(2\hat c_3)^{1/3}}\left[c_0r^6-\frac{\mu}{r^{n-5}}+\hat c_3\tilde J_W-\sqrt{\left(c_0r^6-\frac{\mu}{r^{n-5}}+\hat c_3\tilde J_W\right)^2+4\hat c_3^2\tilde I_W^6}\right]^{1/3} , \label{f_cubic}
\eeq
where $\mu$ is an integration constant and we have defined $I_W^2$ as in~\eqref{IW} and 
\beq
 & & \hat c_3=(n+1)n(n-1)(n-2)(n-3)(n-4)c_3 , \\ 
 & & (n-1)(n-2)(n-3)(n-4)(n-5)\tilde J_W=4\tilde C_{ijkl}\tilde C^{klmn}\tilde C_{mn}^{\phantom{mn}ij}+8\tilde C_{ijkl}\tilde C^{mjkn}\tilde C^{i\phantom{mn}l}_{\phantom{i}mn} .
\eeq

The above solution was obtained in \cite{DadPon15_JHEP} for the special case when $\bh$ is a product of two identical spheres (a solution for cubic Lovelock theory including lower order curvature terms was obtained earlier in \cite{FarDeh14}). When $I_W^6=0$ ($\Rightarrow J_W=0$) the base space is of constant curvature and one recovers the solution obtained in \cite{CaiOht06} (see also \cite{Ray15,Buenoetal16}). 

Comments about static black hole solutions in generic Lovelock gravity with a base space not of constant curvature  can be found in \cite{Ray15,OhaNoz15}.

\section*{Acknowledgments}

S.H. was supported through the Research Council of Norway, Toppforsk
grant no. 250367: \emph{Pseudo-Riemannian Geometry and Polynomial Curvature Invariants:
Classification, Characterisation and Applications.}  M.O. was supported by research plan RVO: 67985840 and research grant GA\v CR 19-09659S.

\renewcommand{\thesection}{\Alph{section}}
\setcounter{section}{0}

\renewcommand{\theequation}{{\thesection}\arabic{equation}}


\begin{thebibliography}{10}

\bibitem{HerOrt19}
Hervik S and Ortaggio M 2020 {\em JHEP\/} {\bf 02}  047

\bibitem{Tangherlini63}
Tangherlini F~R 1963 {\em Il Nuovo Cimento\/} {\bf 27} 636

\bibitem{GibWil87}
Gibbons G~W and Wiltshire D~L 1987 {\em Nucl. Phys. {\rm B}\/} {\bf 287} 717

\bibitem{Birmingham99}
Birmingham D 1999 {\em Class. Quantum Grav.\/} {\bf 16} 1197

\bibitem{DotGle05}
Dotti G and Gleiser R~J 2005 {\em Phys. Lett. {\rm B}\/} {\bf 627} 174

\bibitem{FarDeh14}
Farhangkhah N and Dehghani M~H 2014 {\em Phys. Rev. {\rm D}\/} {\bf 90} 044014

\bibitem{Ray15}
Ray S 2015 {\em Class. Quantum Grav.\/} {\bf 32} 195022

\bibitem{OhaNoz15}
Ohashi S and Nozawa M 2015 {\em Phys. Rev. {\rm D}\/} {\bf 92} 064020

\bibitem{Bleecker79}
Bleecker D~D 1979 {\em J. Diff. Geom.\/} {\bf 14} 599

\bibitem{Wolf68}
Wolf J~A 1968 {\em Acta Math.\/} {\bf 120} 59

\bibitem{Bessebook}
Besse A~L 1987 {\em Einstein Manifolds\/} (Berlin: Springer-Verlag)

\bibitem{Eddington_book}
Eddington A 1930 {\em The Mathematical Theory of Relativity\/} 2nd edn
  (Cambridge: Cambridge University Press)

\bibitem{IyeWal94}
Iyer V and Wald R~M 1994 {\em Phys. Rev. {\rm D}\/} {\bf 50} 846

\bibitem{Gurses92}
G{\"u}rses M 1992 {\em Phys. Rev. {\rm D}\/} {\bf 46} 2522

\bibitem{DotOliTro09}
Dotti G, Oliva J and Troncoso R 2009 {\em Int. J. Mod. Phys. A\/} {\bf 24} 1690

\bibitem{Bogdanosetal09}
Bogdanos C, Charmousis C, Gout\'eraux B and Zegers R 2009 {\em JHEP\/} {\bf 10}
  037

\bibitem{Maeda10}
Maeda H 2010 {\em Phys. Rev. {\rm D}\/} {\bf 81} 124007

\bibitem{DotOliTro10}
Dotti G, Oliva J and Troncoso R 2010 {\em Phys. Rev. {\rm D}\/} {\bf 82} 024002

\bibitem{BouDes85}
Boulware D~G and Deser S 1985 {\em Phys. Rev. Lett.\/} {\bf 55} 2656

\bibitem{Wheeler86_GB}
Wheeler J~T 1986 {\em Nucl. Phys. {\rm B}\/} {\bf 268} 737

\bibitem{Cai02}
Cai R~G 2002 {\em Phys. Rev. {\rm D}\/} {\bf 65} 084014

\bibitem{Lovelock71}
Lovelock D 1971 {\em J. Math. Phys.\/} {\bf 12} 498

\bibitem{DadPon15_JHEP}
Dadhich N and Pons J~M 2015 {\em JHEP\/} {\bf 05} 067

\bibitem{CaiOht06}
Cai R~G and Ohta N 2006 {\em Phys. Rev. {\rm D}\/} {\bf 74} 064001

\bibitem{Buenoetal16}
Bueno P, Cano P~A, Lasso~A. {\'O} and Ram\'{\i}rez P~F 2016 {\em JHEP\/} {\bf
  04} 028

\end{thebibliography}
\end{document}